\documentclass{elsart3p}

 \usepackage{graphics}
 \usepackage{graphicx}
 \usepackage{epsfig}

\usepackage{amssymb}
\usepackage{natbib}
\usepackage{lineno}

\begin{document}

\begin{frontmatter}

\title{Status and prospects of the IceCube neutrino telescope}


\author[CPPM]{Elisa Resconi}{$^*$, for the IceCube Collaboration}

\address[CPPM]{Max-Planck-Institut f\"ur Kernphysik, Saupfercheckweg 1, 69117 Heidelberg, Germany}

\corauth[cor1]{corresponding author: \tt Tel.:+49-6221-516833 \rm \it E-mail address:
\rm elisa.resconi@mpi-hd.mpg.de}

\begin{abstract}
The IceCube neutrino observatory, under construction at the South Pole,
consists of three sub-detectors: a km-scale array of digital optical 
modules deployed deep in the ice,
the AMANDA neutrino telescope and the surface array IceTop.   
We summarize results from searches for cosmic neutrinos with the AMANDA
telescope and review expected sensitivities for IceCube at various installation
phases. Reliability and robustness of installation at the South Pole has been
demonstrated during the past four successful construction seasons. 
The 40 installed IceCube strings are working well. We are developing detailed plans for the final construction of
IceCube, including extensions optimized for low and high energy. 
We describe the IceCube Deep Core
project which will extend the low energy response of IceCube.
\end{abstract}

\end{frontmatter}

\section{Introduction}
\label{sec:intro}

IceCube, the largest neutrino telescope in history is rapidly becoming  a reality after
four successful construction seasons.
IceCube searches primarily for neutrinos in the GeV-PeV energy region 
expected to be
 produced at acceleration sites of high energy cosmic rays \cite{physics} \cite{ic_sensi}. 
The IceCube design \cite{ic_design} employs 80 uniformly spaced strings, each one equipped with 60 digital optical modules (DOMs)
\cite{dom}.
Among the strings of this "in-ice" array, 40 have already been 
deployed at depths between 1.5 and 2.5~km in the icecap of
the South Pole. Moreover, an additional installation in the center-bottom 
part of the detector of 
an inner core array (the so-called IceCube Deep Core)
composed of additional strings has been financed and recently approved. 
The AMANDA-II neutrino telescope \cite{amanda} - the predecessor of IceCube - is maintained in operation. 
AMANDA-II is completely embedded in the IceCube
instrumented volume, and the hits registered by AMANDA and IceCube are merged
together into combined events \cite{ag}.
The third fundamental component of the neutrino observatory is also under construction: 
 the IceTop air shower detector \cite{icetop}. Combining IceTop data with data from the deep-ice
array provides a unique opportunity to study cosmic ray composition in the region of the "knee"
extending earlier measurements performed using the combination of the SPASE and AMANDA
detectors \cite{icetop_p} \cite{Bai:2007pu}.
The PMT pulses of the IceCube DOMs are first converted into digital waveforms.
 Every time the DOM  triggers,
the waveform is digitized, read out and time stamped. 
The digitized PMT pulses from different DOMs are 
sorted into a time-ordered stream. The trigger is accomplished entirely via software. 
Single majority triggering (SMT) is applied in the in-ice array and in the IceTop array. 
For the in-ice, it requires the coincidence of hits in eight or more DOMs within 
a time window of $5 \mu$sec and for IceTop six or more DOMs hit in a time window of the same
size.
Details about the data acquisition can be found in \cite{kael}.
There are two primary detection channels considered for the study of high energy neutrinos:
{\it track-like events} from upward through-going muons induced by charged-current $\nu_{\mu}$ interactions and
 {\it cascade-like events} from charged-current (CC) interactions of $\nu_e$, $\nu_{\tau}$ 
 and neutral-current interactions
 of neutrinos of all flavor. A third channel concerns 
 composite events  in which tracks and cascades from
  high energy CC interactions are observed together \cite{tau}.
Three {\it observables} are available for each detection channel: 
the time of an event which is known
with a relative time resolution \cite{time_res}, the incoming direction characterized by the angular resolution \cite{ic_sensi}
and finally the estimated energy of the event characterized by the energy resolution. Average
values of these observables for the track-like and cascade-like detection channels are reported in Table~\ref{obs}.

\begin{table}
\begin{tabular*}{0.45\textwidth}{@{\extracolsep{\fill}}  l|cc  }
\hline\hline
Track-like Events & IceCube & AMANDA \\  
\hline
Rel. Time Resolution [nsec] & 2 & 5-7\\
Angular Resolution  & $<1^\circ$ & $2^\circ-3^\circ$\\
Energy Resolution (log$_{10}E$) & 0.3-0.4 & 0.3-0.4 \\
Field of view & $2\pi$  & $2\pi$\\
\hline
Cascade-like Events & &\\
\hline
Rel. Time Resolution [nsec] & 2 & 5-7\\
Energy Resolution ($log_{10}E$)& 0.18 & 0.18\\
Field of view & $4\pi$ & $4\pi$\\
\hline\hline
\end{tabular*}
\caption{Best values obtained for AMANDA and expected for IceCube observables.}  
\label{obs} 
\end{table}

The first level processing sorts events by detection channel or by specific
low level analysis channel. Successive higher level processing are progressively more
specific for various physics analyses. For example, track-like events are selected for the search of neutrinos 
from point-like sources \cite{ps_5y}, 
from Gamma-Ray-Bursts (GRBs) \cite{grb_mu}, from annihilation of WIMPs in the Sun \cite{wimp-mu} and 
from a diffuse flux \cite{diffuse_mu}. 
Cascades-like events  are considered for the search of a diffuse neutrino flux \cite{diffuse-oxana} 
and as an additional channel for neutrinos from GRB explosions \cite{grb_diffuse}. 
The data taking of the IceCube observatory started in 2006 with 9 strings (IC-9) 
and continued in 2007 with 22 strings (IC-22). The new physics run with 40 strings started in April 2008 and will 
continue until March 2009. The data taking demonstrated excellent stability and
for the first time, operated successfully during drilling operations. 
As a result, the overall live time of IC-22 has been maximized.
The procedures for installation, drilling operation and detector operation
have been consolidated during the four deployment seasons,
delivering a stable detector that performs as expected or even better without significant problems.

\section{The Past: AMANDA}
The idea to take advantage of  the clear optical characteristics
of the Antarctic ice for a neutrino telescope goes back to
 1990 \cite{b-f} \cite{1990-2}. The first demonstration of the possibility
to detect the Cherenkov light generated by
muons produced by cosmic neutrinos was obtained with the prototype string
deployed in January 1992 at the South Pole \cite{a1}.
In parallel to the progressive understanding of experimental issues like 
optical properties of the deep ice \cite{Askebjer:1994yn}, 
optical coupling between optical modules and refrozen ice, stability of the
equipment during refreezing, drilling of vertical and deep holes etc.,
the study of the physics of non-thermal sources demonstrated the need
of an effective area of the order of 1~ km$^2$ since the very
early time \cite{ic1}.
After the installation of 4 shallow strings (AMANDA-A) \cite{Mock:1995ct} \cite{Askebjer:1995vm} \cite{Hulth:1996tj}
and the observation of the scattering effect produced by air bubbles, 
the installation of ten strings at deeper depth (AMANDA-B10) was completed in 1997 \cite{Hill:1999gk}.
First performance of the 10 strings array demonstrated that the site is adequate for a neutrino
telescope \cite{Wiebusch:1997sq}. The South Pole seasons 1998-1999 and 2000 saw the final
installation of AMANDA-II, which is still taking data in its final configuration (19 strings).
We report here a limited selection of results obtained
during more than 10 years of AMANDA operation.

\subsection{The Point Source Search}
\begin{figure}[t]
\begin{center}
\epsfig{file=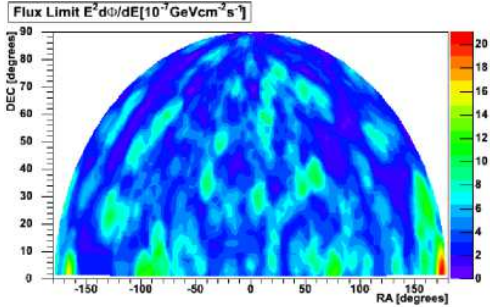 , width=0.40\textwidth}
\end{center} 
\caption{Limit on the extraterrestrial neutrino flux for different position in the sky based on
 AMANDA data collected during 1997-1999 \cite{patrick}.}
\label{ps_97-99}
\end{figure}

\begin{figure}[t]
\begin{center}
\epsfig{file=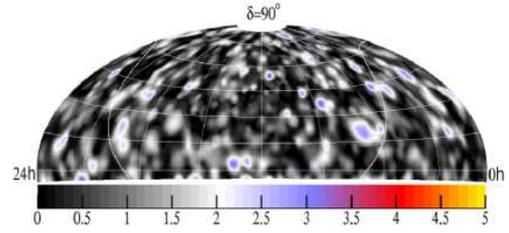 , width=0.40\textwidth}
\end{center} 
\caption{"Sky map" (significance in number of sigmas) of AMANDA data collected during 2000-2006. 
The maximum significance obtained is $3.38 \sigma$ near 11.4h, $+54^\circ$. 
Out of 100 sets of data randomized in right ascension, 
95 have a maximum significance equal to or greater than $3.38 \sigma$. }
\label{ps00-06}
\end{figure}

Downward-going atmospheric muons are the primary background 
 in a high energy neutrino telescope. These are produced in the upper part 
of the atmosphere from the decay of charged pions and kaons which are
generated in the interactions of high energy cosmic rays with atmospheric nuclei.
Highly energetic muons can penetrate a few kilometers of matter. As a result, 
downward-going atmospheric muons 
arrive at the detector site at high rates.
To eliminate the contamination from atmospheric muons, 
neutrino telescopes typically focus on neutrino-induced upward or horizontal-going muons \cite{first}.  
This limits the field of view of a neutrino telescope to 
half of the sky i.e. the opposite hemisphere with respect to the geographical position
of the detector. 
Moreover, atmospheric neutrinos can also induce upward-going muons which are an irreducible
background for extra-terrestrial neutrino searches.
The search for point sources is then realized by looking for localized
excesses over the near-isotropic background of atmospheric neutrinos.
Candidate sources are selected a priori, like active galactic nuclei or Supernova remnants.
The full northern sky is also scanned for unknown sources. Results are reported in Table ~\ref{ps} and
Fig.~\ref{ps_97-99}, Fig.~\ref{ps00-06}.
Methods have been implemented in order to search for variable sources and coincidences 
with X- and $\gamma$-ray flares \cite{Satalecka:2007zz}, \cite{Resconi:2007nx}.

\begin{table}
\begin{tabular*}{0.45\textwidth}{@{\extracolsep{\fill}}  cccc  }
\hline\hline
Detector & Energy & Live time & $E^2_{\nu} \frac{d\Phi_{\nu_\mu}}{dE}$ \\  
Years    &  (TeV) & (days) & (TeV$^{-1}$cm$^{-2}$s$^{-1}$)\\
\hline
AMANDA~B-10 & 1-1000 & 623 & $4.0 \cdot 10^{-10}$ \\
(1997-1999) &&&\\
\hline
AMANDA~II  & 1.6-2600 & 1001 & $5.5 \cdot 10^{-11}$ \\
(2000-2004) &&&\\
\hline         
AMANDA~II  &  & 1387  &  $2.0 \cdot 10^{-11}$ \\
(2000-2006) &&&\\
\hline\hline
IC~22 & 5 - 5000 &270 & $ 1.0 \cdot 10^{-11} $\\
IC~22+AMANDA & 0.1 - 10 & & specific scenarios \\
\hline
IC~80 & 1 - 5000 & 3 years & $2.0 \cdot 10^{-12}$ \\
\hline\hline
\end{tabular*}
\caption{Summary of the best limits achieved for the point source 
search using AMANDA data and expected sensitivity for IceCube analysis. 
The energy range corresponds to that for 90\% of the events, assuming an $E^{-2}$ spectrum.
Details about AMANDA B-10 analysis for the 3 years 97-98-99 are reported 
in \cite{patrick}. The 5 years analysis of AMANDA II is described in \cite{ps_5y} \cite{markus} 
and the latest 
analysis of the 7 years of AMANDA II is reported in \cite{Braun:2007zzb}.}  
\label{ps} 
\end{table}

\subsection{The Diffuse Search}
\begin{figure}[t]
\begin{center}
\epsfig{file=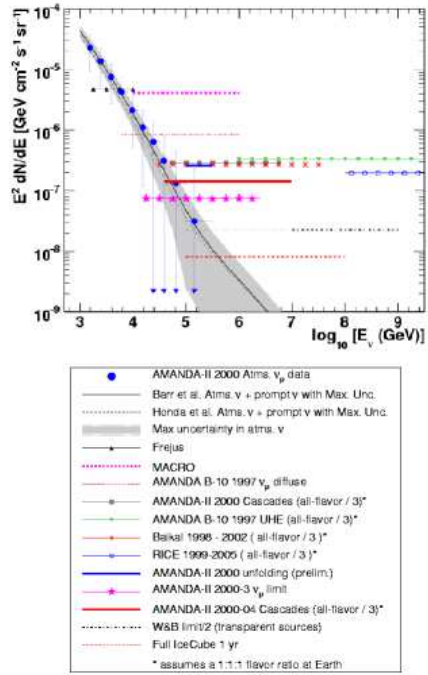, width=0.35\textwidth}
\end{center} 
\caption{Summary of existing experimental limits on the diffuse neutrino flux versus
the logarithm of neutrino energy \cite{diffuse_mu}.}
\label{diffuse}
\end{figure}

Since the very early time of neutrino astronomy, 
the existence of a diffuse flux of neutrinos coming from the sum of faint active 
galaxies has been discussed \cite{Mannheim:1998wp}.
Limits on the diffuse flux have been reported
by the Frejus  experiment \cite{Rhode:1996es} and the Baikal experiment 
\cite{Balkanov:2000in}. Up-going neutrinos have been detected with AMANDA B-10
and AMANDA-II. Latest results using AMANDA data are reported in Fig.~\ref{diffuse}
and in \cite{Achterberg:2007qp}.

\section{The Present: 22 IceCube Strings and AMANDA}
\begin{figure}[t]
\begin{center}
\epsfig{file=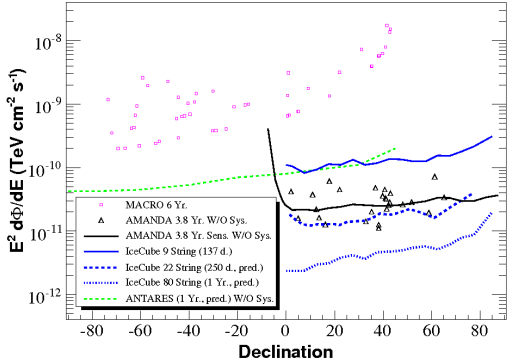 , width=0.45\textwidth}
\end{center} 
\caption{Average (over right ascension) sensitivities vs declination for all sky point 
source searches and upper limits for specific sources. $E^{-2}$ differential spectra are assumed. For references see
\cite{Teresa}.}
\label{ps_sensi}
\end{figure}

In its 2007 configuration, IceCube consisted of 22 strings in operation with 60 digital 
optical modules each. Details of the performance of the 22-strings array
 are reported in \cite{karle}. The sensitivity to point sources provided by the IC22
 array is reported in Fig.~\ref{ps_sensi}. We discuss here in 
 particular the use of AMANDA as an integrated compact core in IceCube.

\subsection{AMANDA: The First Generation Low Energy Core}
\begin{figure}[t]
\begin{center}
\epsfig{file=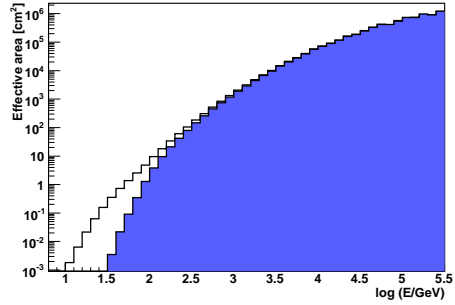, width=0.40\textwidth}
\end{center} 
\caption{Monte Carlo estimation of the muon neutrino effective area at high filter level (L3). In blue the effective area 
of IC22 is reported and in white the one for IC22 and AMANDA. 
}
\label{area}
\end{figure}

\begin{figure}[h]
\begin{center}
\epsfig{file=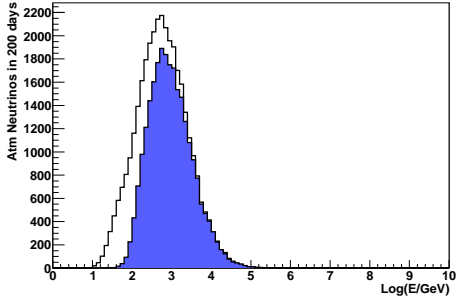, width=0.40\textwidth}
\end{center} 
\caption{Monte Carlo estimation of the atmospheric muon neutrino event rates per 200 days exposure at high filter level (L3). 
 With IceCube-22 alone, 25000 events are observed. With AMANDA included the total rate rises to 34000.}
\label{atm_nu}
\end{figure}

\begin{figure}[h]
\begin{center}
\epsfig{file=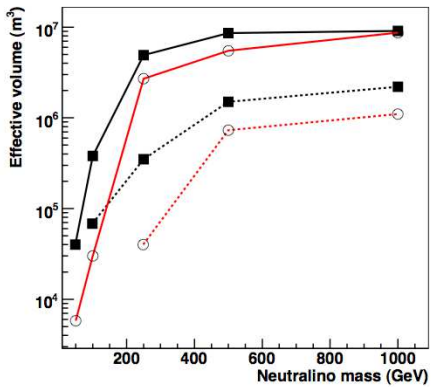, width=0.40\textwidth}
\end{center} 
\caption{
Effective volumes as function of neutralino mass for interacting neutrinos from neutralino annihilation 
in the Sun, for IceCube-22 alone (circles) compared with the combined detector of IceCube-22 
plus AMANDA (squares).  Solid and dashed lines correspond to hard and soft annihilation 
channels, respectively \cite{g1}.}
\label{g1}
\end{figure}

During the two South Pole seasons 2003-2004 and 2004-2005, 
the data acquisition system (DAQ) of AMANDA was significantly upgraded
to provide nearly dead-timeless operation and full digitization 
of the electronic readout \cite{wolfgang}.
This was achieved by using Transient Waveform Recorders (TWR).
The new DAQ system allowed the reduction of the multiplicity trigger threshold 
and, consequently, of the energy threshold below 50 GeV. 
Hence AMANDA was seen to be a complement to
IceCube at low energies and worth a full integration into IceCube.
During 2006-2007 South Pole season, the following items
were deployed in order to realize an integrated AMANDA-IceCube
detector: a common run control unit, triggering and
event building systems, and on-line filtering algorithms \cite{ag}.
Every time the AMANDA detector is triggered, a readout request is sent to the IceCube
detector. The Joint Event Builder (JEB) merges the events
coming from both detectors and provides the data to the on-line filtering.
The on-line filters process the joint events and filter out the interesting ones
which are then transferred via satellite to the Northern Hemisphere for
physics analysis.
The study of AMANDA as a nested array in IceCube reveals that:
\begin{itemize}
\item  because of the additional hits in adjacent IceCube strings,
a larger fraction of AMANDA triggered events can be better reconstructed
and survive at higher analysis level as compared to AMANDA alone;
\item  the 1~km long IceCube strings provide 
a longer lever arm that translates directly into an
improved angular resolution of AMANDA events;
\item the combined detector shows an increased effective area at low energy for 
events below 1~TeV 
down to 10-50~GeV as compared to IceCube alone (Fig.~\ref{area}, Fig.~\ref{atm_nu}).
\end{itemize}
These benefits are used in order to improve the search for
neutrinos from point sources  
and from dark matter annihilation in the Sun. Effective volumes for WIMPs 
are reported in Fig.~\ref{g1} and expected sensitivities in \cite{g1}. Moreover, studies of
atmospheric neutrinos with higher statistics are also conceivable. 
In Section 4, we describe in detail the physics content at lower energies.

The use of AMANDA as a low energy core is limited by
different factors:
\begin{itemize}
 \item
 the position of AMANDA is not optimal 
  inside the IceCube array, being in the higher part of the instrumented 
  volume and in a corner of the array;
 \item
the AMANDA electronics is housed in the MAPO building at the South Pole 
  and the snow accumulation is a threat to the building;
 \item
 the AMANDA electronics is different from IceCube electronics and needs 
  special software and service. 
\end{itemize}
For these reasons, the IceCube collaboration is planning to replace AMANDA with a sub-array
composed by 
additional strings in the center-bottom part of the detector: the IceCube Deep Core.

\section{The Future: IceCube 80, On-line, Deep Core, High Energy Extension}
The potential of the 80 string array is described in \cite{ic_sensi}. Here we summarize
methods for the exploitation of IceCube physics program.  As soon as IceCube reaches
 the sensitivity level to detect short bursts of neutrinos from AGN flares, 
GRB or Supernovae the possibility to send timely alerts to telescopes or satellites
will be of primary importance. 
This requires the development of on-line
analysis, prompt alerting systems and the establishment of collaboration among IceCube and 
photon-based telescopes. Details can be found in \cite{NToT} and in \cite{grb-followup}. 
The IceCube geometry has been 
optimized for the neutrino 
region around 10~TeV. It is desirable, however to improve the performance
at lower energies and higher energies. 
 At higher energies, we are looking into extensions of
IceCube using radio or acoustic technology and also investigating the optimal position of
the remaining IceCube strings. To detect
neutrino fluxes with EeV energies, such as the
guaranteed GZK neutrino flux, a fiducial volume of
approximately 100~km$^3$ is required. 
Optical Cherenkov technology does not allow
the construction of such an array at a reasonable cost, owing to the
relatively short attenuation length of light. 
However, the absorption length of radio and acoustic waves in
the ice may be as long as 1~km, so a sparse array of the
required scale might be feasible.
At the lowest energies, the funding for the IceCube Deep Core sub-array has been 
approved. We report in some details the status about the Deep Core design.

\subsection{The IceCube Deep Core}
\begin{figure}[]
\begin{center}
\epsfig{file=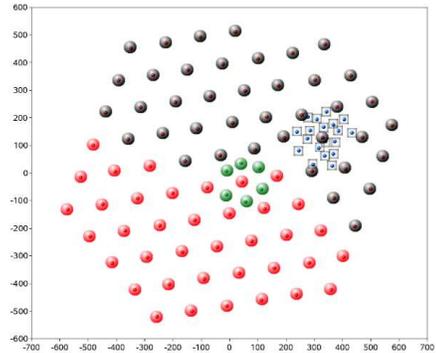, width=0.35\textwidth}
\end{center} 
\caption{Overhead view of the in-ice detector. The black spots represent the
40 strings already deployed. The squares mark the positions of AMANDA strings. 
The red and green spots correspond to the planned IceCube and Deep Core string positions.}
\label{A_IC}
\end{figure}

\begin{figure}[]
\begin{center}
\epsfig{file=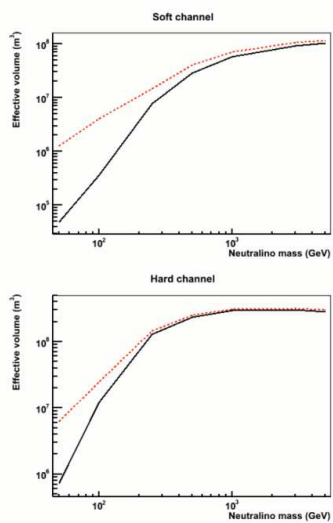, width=0.30 \textwidth}
\end{center} 
\caption{Effective volume as function of neutralino mass 
for muons from interactions of $\nu_{\mu}$ from neutralino 
annihilation in the Sun during the six months it is below the horizon. 
The dashed line corresponds to full IceCube augmented by the six Deep Core strings.
 The full-drawn line corresponds to full IceCube without a deep core array.}
\label{dm}
\end{figure}

\begin{figure}[]
\begin{center}
\epsfig{file=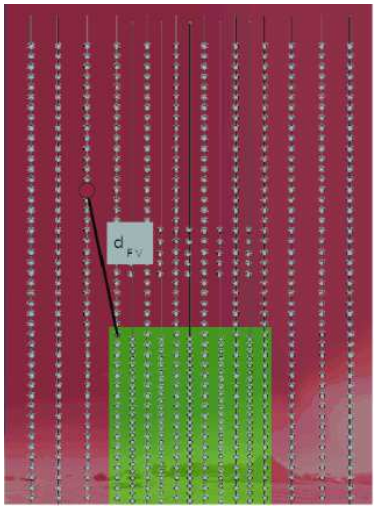 , width=0.30 \textwidth}
\end{center} 
\caption{Schematic view of the IceCube neutrino observatory and the proposed deep core array. 
The Veto Volume (red) and Fiducial Volume(green) are also shown. The distance $d_{FV}$ of 
the veto hits from the FV is used as parameter to separate the cosmic ray muon population from the
neutrino-induced muons}
\label{veto}
\end{figure}

A dedicated IceCube effort towards an improved sensitivity at low energy, the so called IceCube
Deep Core (IC-DC) project, was proposed for the first time in spring 2007.
To have a denser array in the clean ice at large depth, 
below the center of the IceCube array has several advantages compared with AMANDA:
\begin{itemize}
\item
{\it Natural shield}: the larger overburden reduces the background of atmospheric muons.
\item
{\it Muon veto}: the central location allows the use of outer rings of IceCube strings,
 as well as all of the instrumented ice between 1450~m and 2000~m, 
 to achieve additional atmospheric muon veto allowing observations 
 of neutrinos from above the horizon. 
\item
{\it Ice properties}: the larger transparency of the ice at the large depths implies less
 scattering of Cherenkov photons and hence better reconstruction efficiency and better
  angular resolution.
\end{itemize}
At the moment, the baseline for IC-DC calls for six strings centered around one 
 of the central IceCube strings (Fig.~\ref{A_IC}). The spacing among the IC-DC
 strings and the seven closest IceCube strings is 72~m as opposed to the 125~m standard 
 IceCube string spacing. 
 Each string
 will be equipped with 60 IceCube DOMs (Fig.~\ref{veto}).
 Below the so called "dust layer", an ice layer
  at 2100~m with reduced optical transparency,  50 DOMs will be deployed in the very transparent ice 
  with 7.0~m spacing  (IceCube standard is 17~m spacing) and above the dust layer an  
  additional 10 DOMs will be deployed in the transparent ice with 10 m spacing. 
IceCube is optimized for the search of cosmic ray sources able to accelerate up to PeV energies. This 
translates into neutrino energies of the
order of few TeV. However, for several topics in particle physics and astrophysics, 
the detection of lower energy neutrinos is absolutely crucial. Some
examples are:\\
{\it - Dark matter:} neutrinos from neutralino (mass $m_\chi$)  annihilation in 
the Sun and the Earth are expected to have low energy. 
The muon energy from these neutrino interactions in the detector would 
have mean $E \sim m_\chi/3$ for hard and mean $E \sim m_\chi/6$ for soft
 annihilation channels (Fig.~\ref{dm}). One difficulty in the analysis of neutrinos 
 coming from the Sun is that their incoming direction is very close 
 to the horizon (a maximum of 23$^\circ$ below the horizon at the South Pole) 
 where the background of badly reconstructed atmospheric muons is highest.
 The possibility to observe neutrinos from above the horizon will permit to 
 increase the exposure time for neutrinos from dark matter annihilations up to the entire year. \\
{\it -  High energy galactic sources:}
Muon neutrinos produced in the decay chain
$\pi \rightarrow \mu \nu_{\mu} \rightarrow e \nu_{\mu} \nu_{\mu} \nu_e$  
peak at an energy that is a factor of 2-5 lower than the gamma rays from
 $\pi^o \rightarrow \gamma \gamma$ decays \cite{kelner}. In the scenario in which
 gamma rays are of hadronic origin,  
features such as cutoffs observed in gamma ray spectra would be
 expected in neutrino spectra at around half the energy.
Recently the gamma-ray telescopes HESS, VERITAS and MAGIC have 
observed a population of galactic sources characterized by steep 
spectral indices.  
Moreover, several sources show evidence of an 
exponential cutoff in the source spectrum at energies of 10 TeV or so, 
implying that sensitivity at low energy will be essential for observing these sources.  
The sources recently discovered are concentrated in the region of the inner Galaxy, 
most of them in the Southern Hemisphere, i.e. outside the nominal field of view of IceCube. 
The southern sky is a prime target for observations by Mediterranean telescopes such as ANTARES.  
AMANDA and IceCube searches for point sources have focused on the Northern Hemisphere, 
using the Earth to filter out atmospheric muons.  
With IceCube Deep Core we aim not only to lower 
the threshold of IceCube but also to access the Southern Hemisphere 
using part of IceCube as active muon veto. Later in the text more
details about a first implementation of the veto in IceCube are reported.\\
{\it - Atmospheric neutrinos:}  
The ability of the Deep Core array to push the neutrino energy threshold 
down to $\sim 10~$GeV also confers benefits on the study of atmospheric neutrinos.
Their energy spectrum has been measured by AMANDA for energies above 
a few hundred GeV.
With an instrumented physical volume of roughly 20 megatons, 
simulations show that Deep Core will trigger on about 60,000 
atmospheric-neutrino-induced events per year, using a simple 
majority trigger set at six hits in a $5~\mu s$ window and requiring hits 
on at least three strings.  Roughly 8\% of these events are in the 1-10~GeV
 energy bin, 45\% in the 10-100~GeV bin, and 45\% in the 100~GeV - 1~TeV bin.
The measurement of the energy spectrum and angular distribution 
of atmospheric muon neutrinos in the region 10~GeV to 1~TeV 
 is particularly interesting for the study of the transition 
 from pion decay to kaon decay production and for neutrino
 oscillation study. 
 Another aim of Deep Core 
 is to make  a first measurement of the flux of electron neutrinos 
 at energies overlapping with and extending that of Super-Kamiokande \cite{sk}.\\

IceCube offers for the first time the possibility to enlarge the
 field of view of the neutrino telescope using part of the instrumented volume 
 as {\it active veto}. The performance in this respect has been investigated using Monte-Carlo simulation.
 In order to perform a search for neutrinos from above,  
 muons have to be rejected by a factor of $10^6$.
We have divided the detector in two regions: 
Fiducial Volume (FV) and Veto Volume (VV). The FV is the most 
shielded region of the detector i.e. the central and bottom part of the detector 
where the deep core strings will be deployed. Events with their interaction vertex 
in this region are considered signal events. The VV corresponds to the part 
of the instrumented volume required to reject the bulk of the downward 
atmospheric muons. The top 36 DOMs and three layers of strings around the FV 
are used as VV (Fig. ~\ref{veto}).
In order to separate contained and not contained events, we consider 
that a hit far away from the FV has a very low probability to be 
associated to an event in the FV. In order to use this information, 
we penalize the hits proportionately to their distance ($d_{FV}$) from 
the FV giving them an extra weight. Counting hits in the FV and 
using geometrically penalized hits in the VV, we succeed to reject 
downward going muons by four orders of magnitude, still keeping the neutrino signal.
In a second step, a dedicated algorithm determines the most probable 
interaction vertex of the surviving events \cite{Sebastian}. The {\it x,y,z} distributions and 
likelihood of the reconstructed vertex differ significantly from the neutrino 
induced muons and downward through-going muons leaking in.
The final selection of neutrino induced events is made by requiring that 
the reconstructed vertex of the neutrino interaction is inside the FV. 
This allows the rejection of muons 
by two additional orders of magnitude. 
This two-levels analysis proves that IceCube is an efficient muon veto
and fully contained events which trigger IC-DC can be identified with an efficiency 
between 20\% and 30\%.


\begin{thebibliography}{00}

\bibitem{physics}
  F.~Halzen,
  Eur.\ Phys.\ J.\  C {\bf 46} (2006) 669


\bibitem{ic_sensi}
  J.~Ahrens et al., [IceCube Coll.],
  Astropart.\ Phys.\  {\bf 20} (2004) 507

\bibitem{ic_design}
IceCube preliminary design document, J.Ahrens et al. [IceCube Coll.],
http://icecube.wisc.edu

\bibitem{dom}
  K.~Hanson, O.~Tarasova, [IceCube Coll.],
  Nucl.\ Instrum.\ Meth.\  A {\bf 567}, 214 (2006)

\bibitem{amanda}
  J.~Ahrens et al., [AMANDA Coll.],
  arXiv:astro-ph/0211269


\bibitem{ag}
  A.~Gross, C.~Ha, C.~Rott, M.~Tluczykont, E.~Resconi, T.~DeYoung, G.~Wikstr\"om, [IceCube Coll.],
  "The combined AMANDA and IceCube Neutrino Telescope",
{\it ICRC 2007}, arXiv:0711.0353


\bibitem{icetop}
  T.~Waldenmaier,
  Nucl.\ Instrum.\ Meth.\  A {\bf 588} (2008) 130, arXiv:0802.2540 

\bibitem{icetop_p}
 Ahrens J et. al., 2004 Nucl. Instr. Meth. A{\bf 522}, 347-359

\bibitem{Bai:2007pu}
  X.~Bai, T.~K.~Gaisser,  [IceCube Coll.],
  J.\ Phys.\ Conf.\ Ser.\  {\bf 60} (2007) 327


\bibitem{kael}
  K.~D.~Hanson,  [IceCube Coll.],
  J.\ Phys.\ Conf.\ Ser.\  {\bf 60} (2007) 47


\bibitem{tau}
D. F. Cowen, [IceCube Coll.], 
J.\ Phys.\ Conf.\ Ser.\  {\bf 60} (2007) 227.


\bibitem{ps_5y}
  A.~Achterberg et al.,  [IceCube Coll.],
  Phys.\ Rev.\  D {\bf 75} (2007) 102001
  
\bibitem{grb_mu}
  A.~Achterberg, K.~Hurley,  [IceCube Coll.],
  arXiv:0705.1186 
  
\bibitem{wimp-mu}
  D.~Hubert,  [IceCube Coll.],  
astro-ph/0701333

\bibitem{diffuse_mu}
  K.~Hoshina,  [IceCube Coll.], "Diffuse high-energy neutrino searches in AMANDA-II and IceCube: Results
  and future prospects",
  {\it ICRC 2007}, arXiv:0711.0353 


\bibitem{diffuse-oxana}
  O.~Tarasova, M.~Kowalski, M.~Walter,  [IceCube Coll.],
  "Search for neutrino-induced cascades with AMANDA data taken in 2000-2004",
{\it ICRC 2007}, arXiv:0711.0353

  
\bibitem{grb_diffuse}
A.~Achterberg et al.,  [IceCube Coll.],
 astro-ph/0702265

\bibitem{time_res}      
B.~Christy, A.~Olivas, D.~Hardtke, [IceCube Coll.],
"Exotic Particles Searches with IceCube", {\it ICRC 2007}, 
  arXiv:0711.0353 


\bibitem{b-f}                                                                 
S.~Barwick, F.~Halzen,
  Proc.\ Summ.\ Study on High Energy Physics in the 1990's (Snowmass,CO,1990)
  p. 328

  
\bibitem{1990-2}
  S.~W.~Barwick, F.~Halzen, D.~Lowder, T.~Miller, R.~Morse, P.~B.~Price, A.~Westphal,
 Trends in Astroparticle Physics (Santa Monica, CA, 1990) p 413

\bibitem{a1}
  S.~Barwick et al.,
  AIP Conf.\ Proc.\  {\bf 272} (1993) 1250

\bibitem{Askebjer:1994yn}
  P.~Askebjer et al.,  [AMANDA Coll.],
  Science {\bf 267} (1995) 1147


\bibitem{ic1}
  S.~Barwick, F.~Halzen, D.~Lowder, T.~Miller, R.~Morse, P.~B.~Price and A.~Westphal,
  J.\ Phys.\ G {\bf 18} (1992) 225

\bibitem{Mock:1995ct}
  P.~C.~Mock et al.,  [AMANDA Coll.],
  "Status and capabilities of AMANDA-94",
{\it ICRC 95}

\bibitem{Askebjer:1995vm}
  P.~Askebjer et al.,  [AMANDA Coll.],
  Nucl.\ Phys.\ Proc.\ Suppl.\  {\bf 38} (1995) 287

\bibitem{Hulth:1996tj}
  P.~O.~Hulth et al.,  [AMANDA Coll.],
  arXiv:astro-ph/9612068

\bibitem{Wiebusch:1997sq}
  C.~Wiebusch,  [AMANDA Coll.],
{\it 4th SFB-375 Ringberg Workshop on Neutrino Astrophysics, Germany, 1997}

\bibitem{first}
M.A. Markov and I.M. Zheleznykh, Nucl. Phys. 27 (1961) 385. See also M.A. Markov in Proc. 1960 Annual
     International Conference on High Energy Physics at Rochester (ed. E.C.G. Sudarshan, J.H. Tinlot and A.C.
     Melissinos) (1960)

\bibitem{Satalecka:2007zz}
  K.~Satalecka, E.~Bernardini, M.~Ackermann, M.~Tluczykont,  [IceCube
                  Coll.],
  "Cluster search for neutrino flares from pre-defined directions",
{\it ICRC 2007}, arXiv:0711.0353

\bibitem{Resconi:2007nx}
  E.~Resconi,  [IceCube Coll.],
  J.\ Phys.\ Conf.\ Ser.\  {\bf 60} (2007) 223.
  
\bibitem{Mannheim:1998wp}
  K.~Mannheim, R.~J.~Protheroe and J.~P.~Rachen,
  Phys.\ Rev.\  D {\bf 63} (2001) 023003

\bibitem{Rhode:1996es}
  W.~Rhode et al.,  [Frejus Coll.],
  Astropart.\ Phys.\  {\bf 4} (1996) 217

\bibitem{Balkanov:2000in}
  V.~Balkanov et al.,
  Astropart.\ Phys.\  {\bf 25} (2006) 140




\bibitem{Achterberg:2007qp}
  A.~Achterberg et al.,  [IceCube Coll.],
  Phys.\ Rev.\  D {\bf 76} (2007) 042008


\bibitem{karle}
 A.~Karle,  [IceCube Coll.],
 "IceCube: Construction status and performance results of the 22 string detector",
{\it ICRC 2007}, arXiv:0711.0353


\bibitem{wolfgang}
  W.~Wagner,  [AMANDA Coll.],
 "New capabilities of the AMANDA-II high energy neutrino detector",
{\it ICRC 2003}



\bibitem{Hill:1999gk}
  G.~C.~Hill,  [AMANDA Coll.],
  AIP Conf.\ Proc.\  {\bf 516} (2000) 432


\bibitem{patrick}
P.~Berghaus, [IceCube Coll.], astro-ph/0712.4406

\bibitem{Braun:2007zzb}
  J.~Braun, A.~Karle, T.~Montaruli  [IceCube Coll.],
  "Neutrino point source search strategies for AMANDA-II and results from 2005",
{\it ICRC 2007}, arXiv:0711.0353


\bibitem{markus}
  M.~Ackermann,
PhD thesis, DESY-Zeuthen, Germany

\bibitem{g1}
G. Wikstr\"om, Licenciate thesis, Stockholm University, June 2007

\bibitem{sk}
Y. Ashie et al., [Super-Kamiokande Coll.], Phys. Rev. D {\bf 71} (2005) 112005

\bibitem{Sebastian}
S. Euler, 
Diploma Thesis, RWTH Aachen, Germany

\bibitem{kelner}
S.R. Kelner et al., Phys. Rev. D {\bf 74}, 034018 (2006)

\bibitem{Teresa}
  T.~Montaruli et al., [IceCube Coll.],
TAUP 2007, Sendai, Japan 

\bibitem{NToT}
  M.~Ackermann, E.~Bernardini, N.~Galante, F.~Goebel, M.~Hayashida, 
  K.~Satalecka, M.~Tluczykont, R.M.~Wagner et al.,  [IceCube Coll.], [Magic Coll.]
  "Neutrino Triggered Target of Opportunity (NToO) test run with AMANDA-II and MAGIC",
  {\it ICRC 2007}, arXiv:0709.2640 

\bibitem{grb-followup}
  A.~Kappes, M.~Kowalski, E.~Strahler, I.~Taboada,  [IceCube
                  Coll.], "Detecting GRBs with IceCube and optical follow-up observations",
{\it ICRC 2007}, arXiv:0711.0353

\end{thebibliography}
\end{document}